\documentclass[aps,prb,twocolumn]{revtex4-1}
\usepackage{graphicx}
\usepackage{bm}
\usepackage{amsmath,amsthm,amssymb}
\usepackage{gt13}

\usepackage{color}
\begin{document}

\title{
Current-driven dynamics of coupled domain walls in a synthetic antiferromagnet
}
\author{Henri Saarikoski,$^1$ Hiroshi Kohno,$^2$ Christopher H. Marrows,$^3$ and Gen Tatara$^1$
}  
\affiliation{
$^1$RIKEN Center for Emergent Matter Science (CEMS), 2-1 Hirosawa, Wako, Saitama 351-0198, Japan
}
\affiliation{
$^2$Department of Physics, Nagoya University, Nagoya 464-8602, Japan
}
\affiliation{
$^3$School of Physics and Astronomy, University of Leeds, Leeds LS2 9JT, United Kingdom
}
\date{\today}
\begin {abstract} 
We develop the theory of magnetic  domain wall motion in coupled double-layer
systems where electrons can hop between the layers giving rise to
an antiferromagnetic coupling. We demonstrate that the force from
the interlayer coupling drives the walls and the effect of the extrinsic
pinning is greatly reduced if the domain walls are initially separated.
The threshold current density for metastable spin-aligned configurations is also much lower.
We conclude that the interlayer coupling has a significant effect on domain wall mobility in double-layer systems.
\end{abstract}

\maketitle

\section{Introduction}

Domain walls in ferromagnetic materials involve magnetization reversal
in a thin layer. The thickness of this layer is determined by the magnetic anisotropy
energy and the exchange energy. Domain walls separate areas of different
magnetization orientations and they are controllable using currents which
create a spin torque that drives the wall. Therefore, devices using domain
wall dynamics hold promise for future high speed, high density, and
non-volatile data storage.\cite{Hayashi08}

However,  domain wall motion is restricted by intrinsic and extrinsic pinning effects and  current densities needed to move
domain walls are typically high, of the order of $10^{12}$ A/m$^2$. The intrinsic pinning is due to the hard-axis
magnetic anisotropy.\cite{TK04} The extrinsic pinning involves e.g. defects in individual layers.\cite{Thiaville05}
Theoretically, it was demonstrated that in the adiabatic limit, where the wall is driven solely by the spin-transfer torque,
the wall has to overcome the energy barrier arising from the hard-axis anisotropy energy, and that the wall is intrisically pinned.\cite{TK04} 
This energy barrier involves a threshold current below which a domain wall does not move or motion stalls soon after the current is turned on.
In most cases, the threshold current of the intrinsic pinning is high.\cite{Thiaville05}
The intrinsic pinning effect was observed in a perpendicularly magnetized Co/Ni nanowire with reduced hard-axis anisotropy.\cite{Koyama11}
The threshold current density was $2.5\times 10^{11}$ A/m$^2$ and
it was insensitive to the applied magnetic field which was consistent with
theoretical predictions.\cite{TK04}
Spin relaxation results in a torque orthogonal to the spin-transfer torque, and this torque,
called the non-adiabatic torque, removes the intrinsic pinning effect and the threshold current is reduced. \cite{Zhang04,Thiaville05,TTKSNF06}
The threshold current is then determined by the extrinsic pinning potential and the non-adiabaticity parameter, $\beta$.
In principle, the intrisic pinning effect can be removed by fabricating a wire which has a cross-section of a perfect circle.\cite{PhysRevLett.104.057201}
However, most of the experiments have
been carried out in the regime where extrinsic pinning effects dominate. \cite{boulle11}

For realizing fast domain wall motion and low threshold current density, several experimental attempts have been carried out.
Lepadatu et al. controlled the value of non-adiabaticity parameter, $\beta$, by doping permalloy with vanadium.\cite{Lepadatu10}
They showed that V-doping of 10\%  leads to an increase of $\beta$ by a factor of about two, but the threshold current did not
improve since the spin polarization of the current was reduced by doping.\cite{lepadatu10b}
Trilayer Pt/Co/MgO  structures were studied by Miron et al. \cite{Miron10,Miron11}
They were motivated by the idea that Rashba spin-orbit interaction would emerge in interfaces of layers of insulators and metals in the presence of heavy atoms with strong spin-orbit interaction.
This interaction would realize very efficient wall motion since it acts as a large $\beta$ as
predicted theoretically.\cite{Obata08,Manchon09}
The wall velocity in that trilayer system was 400 m/s, which is two orders of magnitude larger than in single layer systems,
at current density of $3\times 10^{12}$ A/m$^2$ (Ref. \onlinecite{Miron11}).
However, it turned out that the mechanism for fast wall motion was not due to the Rashba interaction.
In fact, systematic analysis on structures Pt/CoFe/MgO and Ta/CoFe/MgO indicated that the spin Hall effect in Pt and Ta layer injects
spin current into the ferromagnetic layer and induces a substantial torque on the domain wall, resulting in fast motion.\cite{Emori13,Ryu13}
Due to high domain wall velocities  artificial multi-layered structures are promising for designing devices with efficient domain wall motion. 

Here we develop theory of domain wall motion in coupled double-layer
systems where electrons can hop between the layers giving rise to
an antiferromagnetic coupling. Presence of antiferromagnetic coupling between
the layers was demonstrated\cite{PhysRevLett.57.2442} for ultrathin films in mid-1980's 
and it is used in applications such as magnetic stabilization of magnetoresistive recording heads.\cite{Pinarbasi99}
The interlayer coupling induces an attractive force between the walls
in the two different layers, and this force is expected to help depin the wall since the current drives both walls.
We derive equations of motion for the system in the presence of force
from the interlayer coupling and calculate domain wall dynamics from the resulting equations.
It turns out that interlayer coupling indeed reduces the threshold current greatly if domain walls are initially separated at different pinning sites.
The coupled layer systems are therefore promising for efficient domain wall motion not affected by localized random defects. 

\section{Theoretical model}

\begin{figure}
\includegraphics[width=\columnwidth]{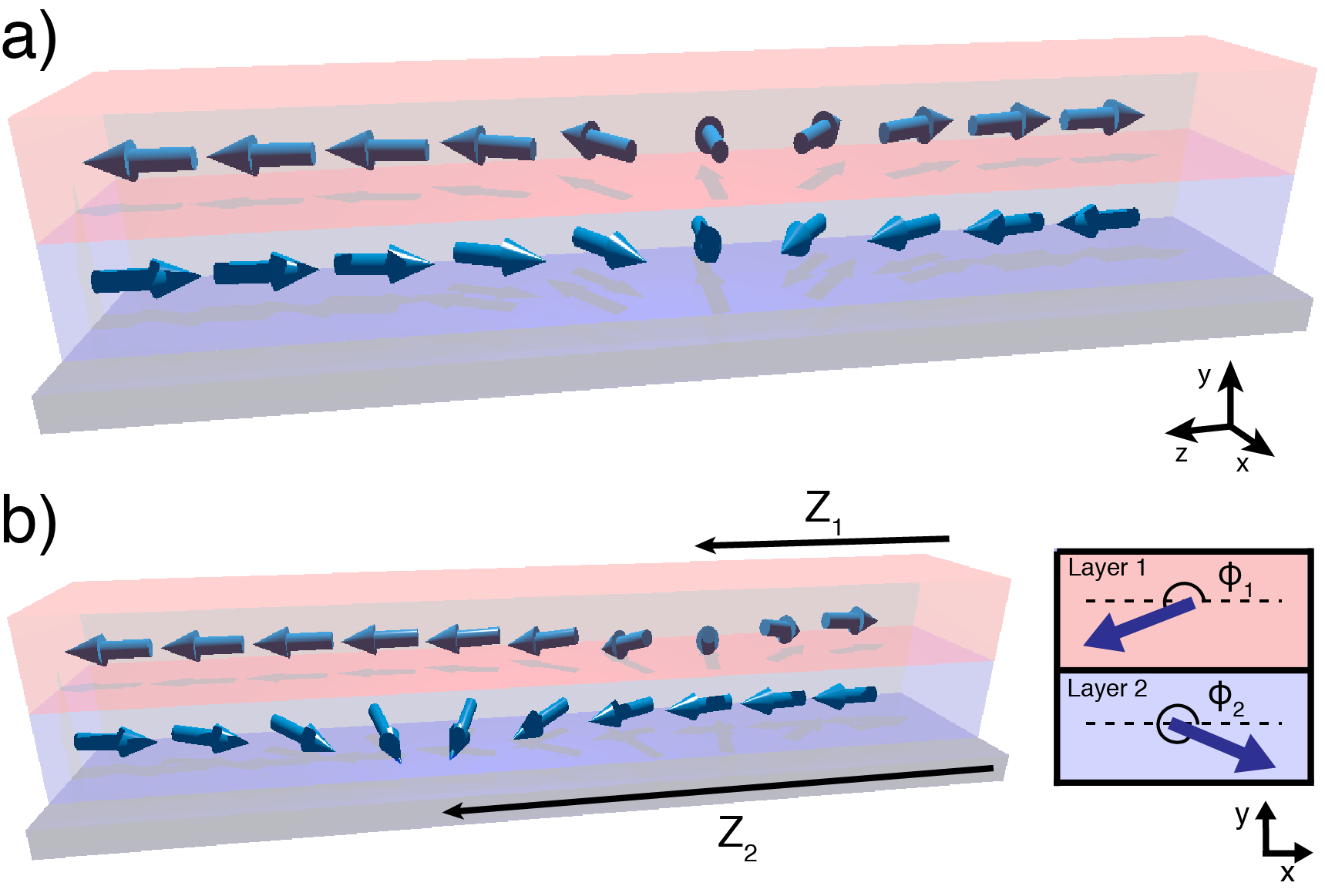}
\caption{a) Ground-state geometry of a N\'eel-type domain walls in a  synthetic double-layer antiferromagnet.
b) Coupling between the layers gives rise to an attractive force between the walls at finite separation $Z_1-Z_2$.
The figure shows a spin configuration which has out-of-plane angles $\phi_{1,2}$. Spin orientation at the center of the 
walls are shown in the inset.
\label{domwall}}
\end{figure}

In this section we derive equation of motion for domain walls in a ferromagnetic double-layer system.
We label the layers by $i=1,2$.
The localized spin direction at position $\textit{\textbf{r}}$ and time $t$ in each ferromagnetic layer is denoted by a unit
vector field, ${\textit{\textbf{n}}}^{(i)}(\textit{\textbf{r}},t)$. We define a coordinate system such that the wire lies in the $x$-$z$
plane, extended along the $z$-direction, with the two layers stacked above each other in the $y$-direction (see Fig. \ref{domwall}a).
The magnetic easy axis is along the $z$-direction and $y$-direction is the magnetic hard axis.
The spin Hamiltonian can then be written as
\begin{align}
H_S= \sum_{i=1,2} \int_{V_i} \frac{d^3r}{a^3} \times  &\nonumber \\ 
\bigg ( \left(\frac{JS^2}{2}\nabla{\textit{\textbf{n}}}^{(i)} \right)^2 -
\frac{KS^2}{2}\left (n_z^{(i)} \right )^2 &
+\frac{K_\perp S^2}{2}\left ( n_y^{(i)} \right )^2 \bigg),
\end{align}
where $J$, $K$ and $K_\perp$ are the strength of the exchange interaction, the easy axis anisotropy energy and the hard axis anisotropy energy,
respectively. In nanowires made from magnetically soft permalloy-like materials, these anisotropy constants arise from shape
anisotropy. The magnitude of spin is $S$, $a$ is a lattice constant and $V_i$ denotes volume of the ferromagnet $i$.
We consider the case in which material constants are the same for both ferromagnets.

Coupling between the two ferromagnets is mediated by electron hopping between the layers.
The in-plane component of the coupling is here antiferromagnetic  $\Delta_\parallel \geq 0$. 
We assume for generality that the out-of-plane component of the interlayer coupling $\Delta_\perp$ is different from
the in-plane component, since this coupling is affected by  the demagnetization  field. Therefore we consider here
both antiferromagnetic $\Delta_\perp>0$ and ferromagnetic $\Delta_\perp<0$ out-of-plane couplings.
We assume that the two ferromagnets are thin (compared with the domain wall thickness) and that the interlayer
coupling acts uniformly on the whole spins. 
The interlayer coupling is thus represented by Hamiltonian
\begin{align}
H_{\rm I} &=   \int_{V_1} 
\frac{d^3r_1}{a^3}\int_{V_2} \frac{d^3r_2}{a^3}
\big[ \Delta_\parallel S^2 (n_x^{(1)}({\textit{\textbf{r}}}_1)n_x^{(2)}({\textit{\textbf{r}}}_2)+ \nonumber \\
& n_z^{(1)}({\textit{\textbf{r}}}_1)n_z^{(2)}({\textit{\textbf{r}}}_2) ) 
  + \Delta_\perp S^2 n_y^{(1)}({\textit{\textbf{r}}}_1)n_y^{(2)}({\textit{\textbf{r}}}_2) \big].
\end{align}

Magnetic anisotropy is very common in thin ferromagnetic films. Therefore we consider only N\'eel-type domain walls which have
the domain wall solution
\begin{align}
{\textit{\textbf{n}}}^{(i)} &= \left( \begin{array}{c} \sin\theta_i\cos\phi_i \\ \sin\theta_i\sin\phi_i \\ \cos\theta_i \end{array} \right),
\end{align}
where  
\begin{align}
\cos\theta_i =(-)^i\tanh \frac{z-Z_i(t)}{\lambda} \label{DWsol}
\end{align}
and $\sin\theta_i=[\cosh \frac{z-Z_i(t)}{\lambda}]^{-1}$, where $\theta$ is the angle between the moment and the $z$-axis
and $\phi_i(t)$ is the azimuthal angle around that axis and represents the out-of-plane angle of the spin in Fig. \ref{domwall}b.
The wall position is denoted by $Z_i$ and $\lambda$ is the thickness of the wall, given by $\lambda=\sqrt{J/K}$.
The topological charge of the domain wall, given by the sign in Eq. (\ref{DWsol}), differs for the two domain walls as a result of the antiferromagnetic
in-plane coupling. This property is essential in the dynamics of the system of coupled walls.
The geometry of the synthetic antiferromagnet under consideration is shown in Fig.~\ref{domwall}.

Current applied in the direction of the wire gives rise to two important effects;
the adiabatic spin-transfer torque effect, which induces a torque on the domain wall,
and a non-adiabatic contribution which is described as a force on the wall.
The adiabatic effect is given by the spin-transfer Hamiltonian,
\begin{align}
H_{\rm ST} &= -\sum_{i=1,2} \int_{V_i} 
\frac{d^3r}{a^3}\hbar S \frac{Pa^3}{2eS}(\textit{\textbf{j}}\cdot\nabla)\phi_i \left(\cos\theta_i-1\right),
\end{align}
where $\textit{\textbf{j}}$ is the electric current density, $e(<0)$ is the electron charge, $P$ is the spin polarization of the current. 
The non-adiabatic contribution as well as damping are inserted later in the equations of motion.

\begin{figure}
\includegraphics[width=\columnwidth]{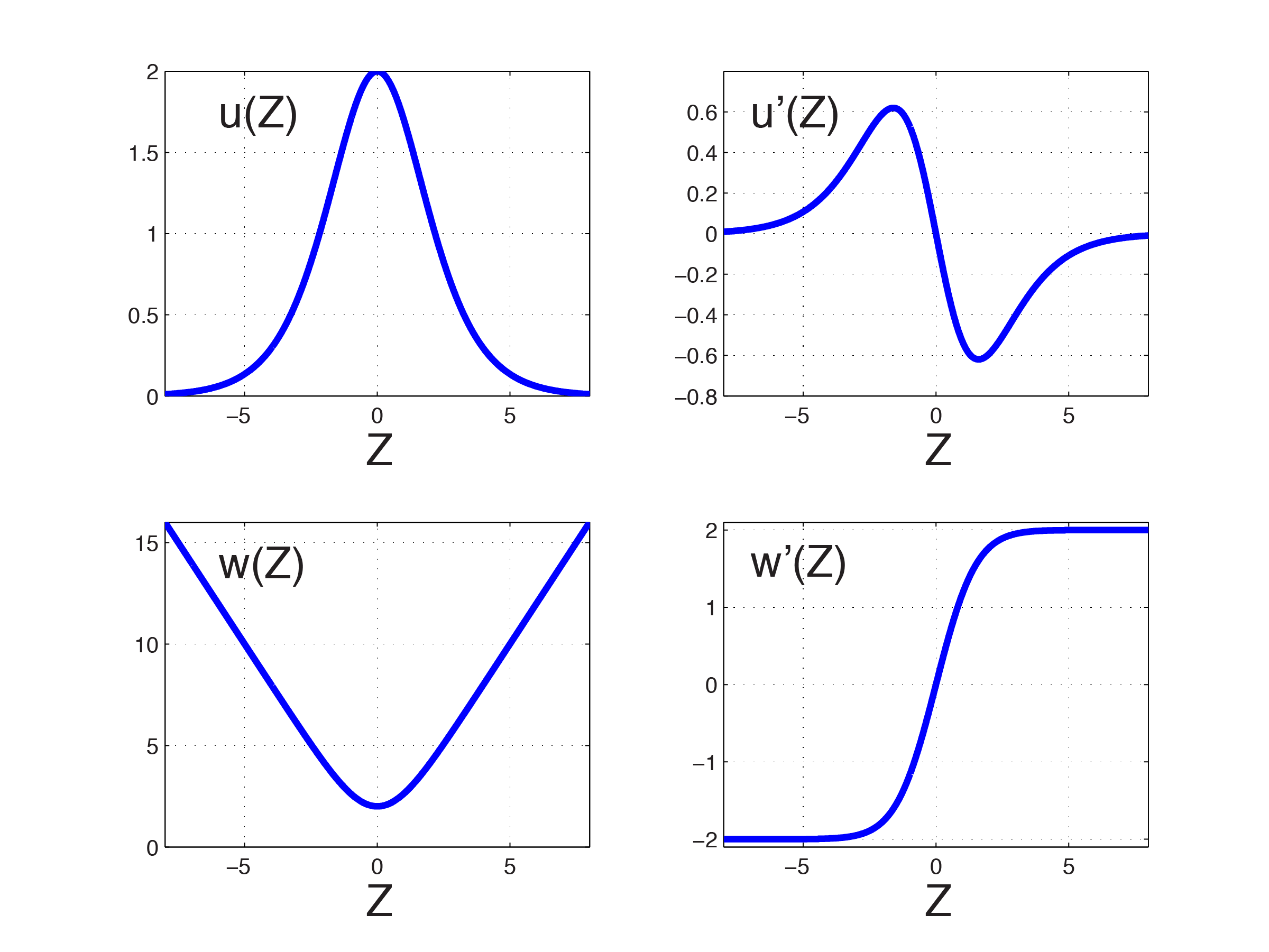}
\caption{ Potentials $u(Z)$ and $w(Z)$ and their derivatives, $u'(Z)$ and $w'(Z)$.  \label{FIGuw}}
\end{figure}

By collecting all the above contributions we obtain the Lagrangian of the coupled double-layer system under applied current 
\begin{align}
L &= \sum_{i=1,2} \int_{V_i} 
\frac{d^3r}{a^3}\hbar S \dot\phi_i \left(\cos\theta_i-1\right)
-H_S-H_{\rm I}-H_{\rm ST},  \label{L}
\end{align}
where the first term is the spin Berry's phase term.
We rewrite the Lagrangian in terms of the collective coordinates for the two walls, $Z_i(t)$ and $\phi_i(t)$ (see Fig.~\ref{domwall}b).
The spin Berry's phase term reduces to
\begin{align}
\sum_{i=1,2} \int_{V_i} 
\frac{d^3r}{a^3}\hbar S \dot\phi_i \left(\cos\theta_i-1\right)
&=
\sum_{i=1,2} \hbar N_i S (-)^i \phi_i \dot {Z}_i,
\end{align}
where $N_i\equiv\frac{2\lambda A_i}{a^3}$ is the number of spins in the wall, $A_i$ is the cross-sectional area of the wire, and
we used the fact that $\dot\phi_i \cos\theta_i$ is equivalent to $-\phi_i \frac{d}{dt}\cos\theta_i $ using integration by parts.
One can easily show that
\begin{align}
H_S
&=
\sum_{i=1,2} N_i S^2 \frac{K_\perp}{2} \sin^2 \phi_i ,
\end{align}
and
\begin{align}
H_{\rm I} &=   N_{\rm I} S 
\big[ \Delta_\parallel u((Z_1-Z_2)/2)\cos\phi_1\cos\phi_2 \nonumber \\
  & + \Delta_\perp u((Z_1-Z_2)/2)\sin\phi_1\sin\phi_2 \nonumber \\
 & + \Delta_\parallel w((Z_1-Z_2)/2)
 \big],
\end{align}
where $N_{\rm I}$ is the effective number of spins and 
\begin{align}
u(Z) &\equiv \int_{-\infty}^{\infty}dz \frac{1}{\cosh z \; \cosh (z-2Z)} \nonumber \\ & = 2 Z {\rm csch}(Z). \\ 
w(Z) &\equiv \int_{-\infty}^{\infty}dz \left(1-\tanh z \; \tanh (z-2Z)\right) \nonumber\\
 &= 2 Z {\rm coth}(Z).
\end{align}
The potentials $u(Z)$, $w(Z)$ and their derivatives are plotted in Fig. \ref{FIGuw}.
After these calculations the Lagrangian reads
\begin{align}
L
&=
\sum_{i=1,2} \hbar N_i S \left[ (-)^i \frac{\phi_i}{\lambda} 
\left(\dot Z_i - \nu_e\right)-\nu_c\sin^2\phi_i\right]  \nonumber\\
& 
-\hbar N_{\rm I} S \big[ \Delta_\parallel u((Z_1-Z_2)/2)\cos\phi_1\cos\phi_2 \nonumber \\
&+ \Delta_\perp u((Z_1-Z_2)/2)\sin\phi_1\sin\phi_2 \nonumber \\
&+ \Delta_\parallel w((Z_1-Z_2)/2)
 \big],
\end{align}
where $\nu_c\equiv \frac{K_\perp \lambda S}{2\hbar}$ and $\nu_e\equiv\frac{Pa^3}{2eS}j$.

Now we include the effect of damping and non-adiabatic contribution of the current.
The damping\cite{TKS_PR08} is included as
$\frac{\delta L}{\delta Z_i}=\alpha \hbar N_i S \frac{\dot{Z}_i}{\lambda}$ and
$\frac{\delta L}{\delta \phi_i}=\alpha \hbar N_i S {\dot{\phi}_i}$.
The non-adiabatic torque, represented by a parameter $\beta$, induces a force 
$\frac{\hbar N_i a^3}{2eS\lambda^2}\beta Pj$.
Since $L=-\hbar N_1 S \frac{\phi_1}{\lambda}\dot Z_1+ \hbar N_2 S \frac{\phi_2}{\lambda}\dot Z_2-H$,  the equations of motion
obtained by differentiating with respect to $Z_1$ and $Z_2$ are 
\begin{align}
 - \hbar N_1 S\frac{\dot{\phi}_1}{\lambda}
&= F_1\nonumber\\
  \hbar N_2 S\frac{\dot{\phi}_1}{\lambda}
&= F_2,
\nonumber\\
\end{align}
where 
\begin{align}
F_1 &\equiv -\frac{\delta H}{\delta Z_1}-\alpha \hbar N_1 S \frac{\dot{Z}_i}{\lambda}+ \frac{\hbar N_1 a^3}{2eS\lambda^2}\beta Pj
\nonumber\\
F_2 &\equiv -\frac{\delta H}{\delta Z_2}-\alpha \hbar N_2 S \frac{\dot{Z}_i}{\lambda}+ \frac{\hbar N_2 a^3}{2eS\lambda^2}\beta Pj
\nonumber\\
\end{align}
are the forces.
The equations of motion thus read
\begin{align}
-\dot{Z}_1 - \alpha\lambda \dot{\phi}_1 
&=
\nu_c \sin 2\phi_1 -\nu_e \nonumber \\
& -\mu_1 u((Z_1-Z_2)/2) \big(\Delta_+ \sin(\phi_1-\phi_2) \nonumber \\
& +\Delta_- \sin(\phi_1+\phi_2) \big)
\label{fulleq1}
\\
\dot{\phi}_1-\alpha\frac{\dot{Z}_1}{\lambda}
&= 
-\frac{\beta}{\lambda} \nu_e +\frac{\mu_1}{2}\bigg[\Delta_\parallel w'((Z_1-Z_2)/2) \nonumber \\
&  +u'((Z_1-Z_2)/2)  \big(\Delta_ +\cos(\phi_1-\phi_2)+ \nonumber \\
&\Delta_-\cos(\phi_1+\phi_2) \big)\bigg]
\label{fulleq2} \\
\dot{Z}_2 -\alpha\lambda \dot{\phi}_2 
&=
\nu_c \sin 2\phi_2 + \nu_e \nonumber \\
 &+\mu_2 u((Z_1-Z_2)/2) \big(\Delta_+ \sin(\phi_1-\phi_2) \nonumber \\
&- \Delta_- \sin(\phi_1+\phi_2) \big)
\label{fulleq3}
\\
-\dot{\phi}_2-\alpha\frac{\dot{Z}_2}{\lambda}
&= 
-\frac{\beta}{\lambda} \nu_e - \frac{\mu_2}{2}\bigg[\Delta_\parallel w'((Z_1-Z_2)/2)  \nonumber \\
&  +u'((Z_1-Z_2)/2)\big(\Delta_+\cos(\phi_1-\phi_2) \nonumber \\ &+\Delta_-\cos(\phi_1+\phi_2) \big)\bigg] ,
\label{fulleq4}
\end{align}
where $\Delta_\pm\equiv \frac{1}{2}(\Delta_\parallel\pm\Delta_\perp)$ and
$\mu_i\equiv N_{\rm I}/N_i$. The $\mu_i$ parameters of the planes determine whether the system is a balanced synthetic antiferromagnet
at $\mu_1=\mu_2$ or an unbalanced synthetic ferrimagnet at $\mu_1 \neq \mu_2$.

\section{Effect of pinning}

\label{pinningpot}
Domain wall dynamics is affected by impurities, notches and other non-uniformities in the layers.
We model such non-uniformities using pinning forces on the domain walls.
We are interested in calculating the terminal velocity of the domain walls under applied current when the domain walls are initially pinned in both layers.
We consider therefore one pinning potential in each layer at distance $\ell$ to each other \cite{TKS_PR08}
\begin{align}
F&= -k_0^{(1)} (Z_1-\ell) \theta(\xi-|Z_1-\ell|) -k_0^{(2)} Z_2 \theta(\xi-|Z_2|),
\end{align}
where $k_0^{(i)}$ ($i=1,2$) are constants representing the strength of the potentials, $\xi$ is the width of the potential and $\theta(x)$ is a step function.
We set the potential width $\xi$ in both layers.
Using $k_i\equiv \frac{\lambda}{\hbar N_i S}k_0^{(i)}$, defining center of mass $Z_+$ and the difference $Z_-$
in the domain wall positions using $Z_\pm\equiv \frac{1}{2}(Z_1 \pm Z_2)$ as well as the 
average phase $\phi_+$ and the difference in the phase using $\phi_\pm\equiv \frac{1}{2}(\phi_1 \pm \phi_2)$, and denoting
$\mu_\pm=(\mu_1 \pm \mu_2)/2$ we obtain the final
equations for motion
\begin{flalign}
&\dot{Z}_+ +\alpha\lambda \dot{\phi}_- 
 =
-\nu_c \cos 2\phi_+\sin2\phi_- +\nu_e &\nonumber \\
& + u(Z_-)  \big(\mu_+\Delta_+ \sin(2\phi_-)+ \mu_-\Delta_- \sin(2\phi_+) \big),
\label{fulleqswithpinning1} 
\end{flalign}
\begin{flalign}
&\dot{\phi}_- -\alpha\frac{\dot{Z}_+}{\lambda} =
\frac{k_1}{2}(Z_++Z_- -\ell) \theta(\xi-|Z_++Z_--\ell|)  \nonumber \\
& +\frac{k_2}{2}(Z_+-Z_-)\theta(\xi-|Z_+-Z_-|) - \frac{\beta}{\lambda} \nu_e +\frac{\mu_-}{2} \times \nonumber \\
&  \big[\Delta_\parallel w'(Z_-)
+u'(Z_-)\left(\Delta_+\cos(2\phi_-)+\Delta_-\cos(2\phi_+) \right)\big],  \nonumber \\
\label{fulleqswithpinning2}
\end{flalign}
\begin{flalign}
&\dot{Z}_- +\alpha\lambda \dot{\phi}_+ 
=
-\nu_c \sin 2\phi_+ \cos 2\phi_- \nonumber \\
&+u(Z_-) \big(\mu_-\Delta_+ \sin(2\phi_-) + \mu_+\Delta_- \sin(2\phi_+) \big),
\label{fulleqswithpinning3}
\end{flalign}
\begin{flalign}
&\dot{\phi}_+ - \alpha\frac{\dot{Z}_-}{\lambda}
= \frac{k_1}{2}(Z_++Z_- -\ell)\theta(\xi-|Z_++Z_--\ell|) \nonumber \\
& -\frac{k_2}{2}(Z_+-Z_-)\theta(\xi-|Z_+-Z_-|) + \frac{\mu_+}{2}\times \nonumber\\
& \left[\Delta_\parallel w'(Z_-)+u'(Z_-)
  \left(\Delta_+\cos(2\phi_-)+\Delta_-\cos(2\phi_+) \right)\right] ,
\label{fulleqswithpinning4}
\end{flalign} 


\section{Terminal velocity of unpinned domain walls}

 \label{terminal}
In the absence of pinning potentials the terminal velocity of the domain wall can be analytically solved.
We first assume that the domain wall separation remains small i.e. $|Z_-|\ll \lambda$ which gives approximately
$u'(Z_-)=0$ and $w'(Z_-)=0$.
We assume also that $\phi_\pm$ changes with time, resulting in vanishing of time averages of $\sin2\phi_\pm$ and $\cos2\phi_\pm$.
After time-averaging we see that the terminal velocities are not affected by the interlayer coupling
\begin{align}
\average{ \dot{Z}_+ }
&=
\frac{1}{1+\alpha^2} \nu_e(1+\alpha\beta) 
\nonumber\\
\average{ \dot{\phi}_- } 
&= 
\frac{1/\lambda}{1+\alpha^2}\nu_e(\alpha-\beta)
\nonumber\\
\average{ \dot{Z}_- }
&=
0
\nonumber\\
\average{ \dot{\phi}_+ } 
&= 
0.
\label{terminal1}
\end{align} 

We then assume that the separation of the domain wall grows with time, e.g., $|Z_-|\gg \lambda$.
We can then approximate $u(Z_-)=u'(Z_-)=0$ and $w'(Z_-)=2\,\mbox{\rm sgn}(Z_-)$.
After time averaging the terminal velocities are then
\begin{align}
\average{ \dot{Z}_+ }
&=
\frac{1}{1+\alpha^2}\left[ \nu_e(1+\alpha\beta)+\alpha \lambda\mu_- \Delta_\parallel \mbox{\rm sgn}(Z_-)\right]
\nonumber\\
\average{ \dot{\phi}_- } 
&= 
\frac{1/\lambda}{1+\alpha^2}\left[ \nu_e(\alpha-\beta)+ \lambda \mu_- \Delta_\parallel \mbox{\rm sgn}(Z_-)\right]
\nonumber\\
\average{ \dot{Z}_- }
&=
- \frac{1}{1+\alpha^2}\alpha \mu_+ \Delta_\parallel \mbox{\rm sgn}(Z_-)
\nonumber\\
\average{ \dot{\phi}_+ } 
&= 
\frac{1/\lambda}{1+\alpha^2} \mu_+ \Delta_\parallel \mbox{\rm sgn}(Z_-).
\label{terminal2}
\end{align} 
We see that in this limit the velocity increases with interlayer coupling. This can be understood from the antiferromagnetic coupling of the walls which
exerts a force on the walls. In practice unpinned walls which are first at finite distance from each other move fast until the separation vanishes. Then the walls
start moving together at a lower velocity determined by Eq. (\ref{terminal1}). This typical behaviour is shown in numerical simulations in
Fig. \ref{dwmotion}.

%

\section{Threshold current}

Eqs. (\ref{fulleqswithpinning1})--(\ref{fulleqswithpinning4}) are a group of first order differential equations. 
We integrate the solution from initial conditions
using a numerical Runge-Kutta-Fehlberg 4th order method with a 5th order error estimator for the adaptive step size. 
We use dimensionless units in calculations by fixing $\nu_c=1$ and setting the thickness of the wall $\lambda=1$.
The time is measured in terms of a dimensionless quantity, 
$t\nu_c/\lambda=tK_\perp S/(2\hbar)$.
We consider separately the  adiabatic and non-adiabatic regimes. In
the former regime adiabatic torque on the wall dominates dynamics
and in the latter case the non-adiabatic force gives the most important contribution.

Due to the terms which depend on $Z_\pm$ and $\phi_\pm$
the time evolution of the system depends on the initial separation of
the domain walls as well as the difference in their phases.
A slight variation in the initial domain wall positions and phases is introduced in order to simulate experimental
situations at finite temperature and in order to avoid special limiting solutions to the differential equations,
for instance when terms on the right hand side vanish at $\phi_+=\phi_-=0$.
This also smooths out the effect of discontinuous external pinning potentials.
We use tiny displacements to the initial domain wall positions and
phases using random numbers from a uniform distribution giving a $\pm 0.01$
change in the wall position with respect to each other (in units of
wall thickness $\lambda$).
The domain wall velocity is evaluated after a sufficiently long time when the domain wall motion has stabilized.

The initial fluctuation of the wall position, $\delta Z$, corresponds to energy fluctuation of 
$\delta {\cal E}=\frac{k_0}{2}(\delta Z)^2=NV_0(\delta Z/\xi)^2$, where the pinning potential depth per spin is 
$V_0\equiv \frac{\hbar S}{2\lambda}k\xi^2$ 
(we suppress here the suffix $i=1,2$ denoting the layer). 
In numerical calculations, the time is measured in terms of a dimensionless quantity, 
$t\nu_c/\lambda=tK_\perp S/(2\hbar)$, and thus a pinning strength $k=0.1$ we use in the calculations
would correspond to the pinning potential of 
$V_0/K_\perp=kS^2\xi^2/(4\nu_c)=2.5\times 10^{-2}$ if we choose $S\simeq1$ and $\xi\simeq\lambda$.
For permalloy wires, $K_\perp\sim 0.03 \sim 2.4\,$K (Ref. \onlinecite{Yamaguchi06}), and if we consider a wall with thickness of 100 nm in a wire of
cross-sectional area of 400 nm $\times 5$ nm, we have $N=1.3\times 10^7$ (for $a=2.5$\,\AA) as the number of spins in the wall.
The fluctuation energy for $\delta Z=0.01$ therefore is
$\delta{\cal E}=30K_\perp=1\sim72\,$ K.
The initial fluctuations of $\pm 0.01$ in the calculations is therefore small in magnitude in comparison to those expected for permalloy wires at
room temperature.

\subsection{Extrinsic pinning}

We insert pinning potentials in both layers as described in Section \ref{pinningpot}.
The initial conditions are chosen to fix the domain walls at the center of  the pinning potentials. 
Details of the domain wall dynamics depend now on the relative strength of the parameters in the model.
At finite values of the non-adiabatic torque $\beta$ the extrinsic pinning
potentials usually restrict domain wall motion and a large driving current is needed to depin the walls. This limits usefulness of magnetic domains in
applications, and means to improve mobility has been in the focus of intense research efforts.

\begin{figure}
\includegraphics[width=\columnwidth]{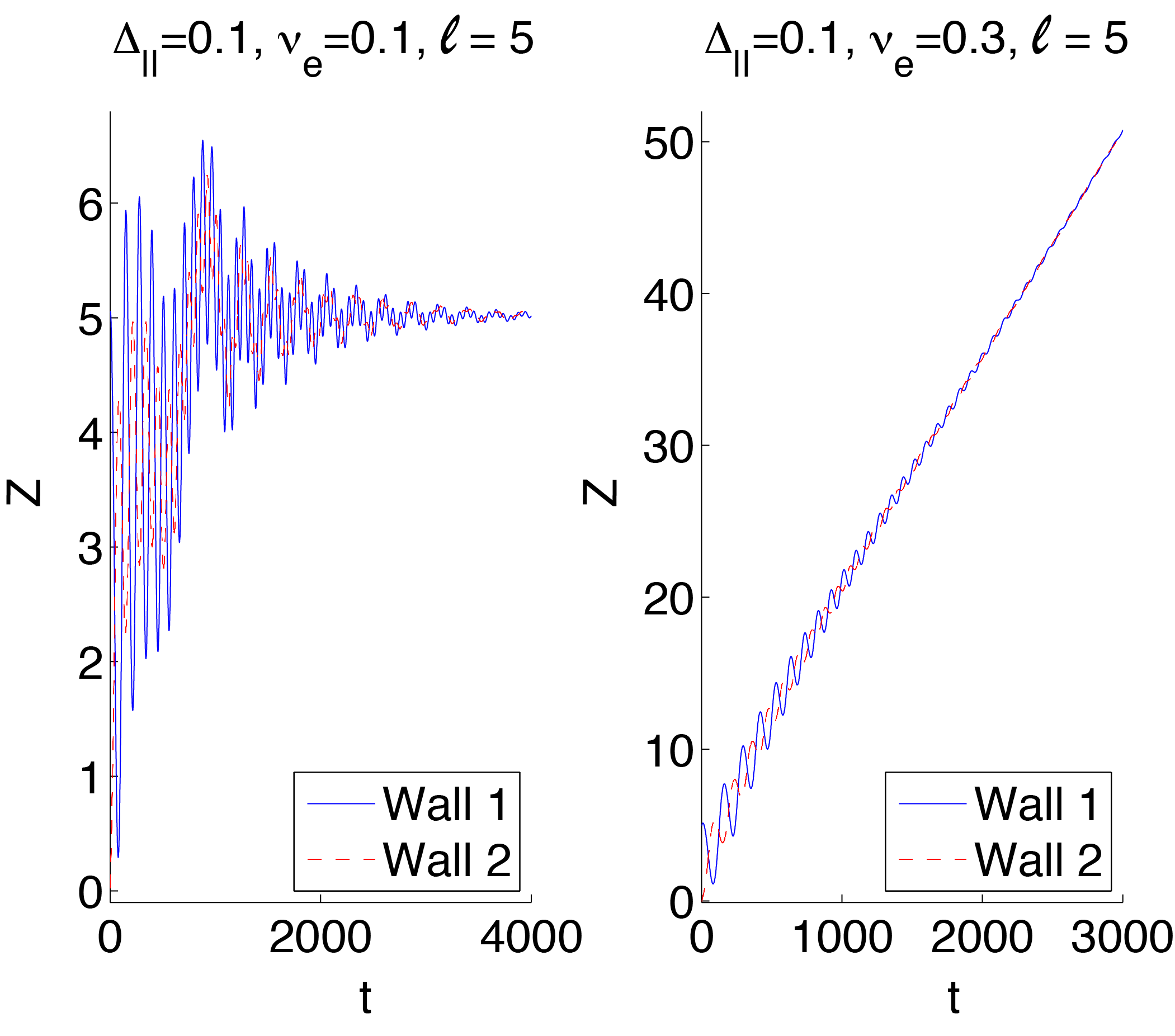}
\caption{Motion of coupled domain walls in double layer systems under extrinsic pinning. The current is switched on at time $t=0$
and positions $Z$ of the two domain walls in the system are calculated
for pinning potentials which are located at $Z_1=5$ and at $Z_2=0$.
When the current is insufficient to unpin the walls they oscillate in the pinning potentials with dampening amplitude (left). 
When the current is sufficiently large, it unpins the walls aided by the force from the interlayer coupling and the walls start moving (right).
Eventually the walls move together with a vanishing phase difference. The terminal velocity is calculated from this limiting motion. \label{dwmotion}}
\end{figure}

Figure \ref{dwmotion} shows typical domain wall dynamics when
the walls are pinned by the potentials and when the interlayer
coupling is large enough to
unpin the walls, respectively. If the force from the interlayer coupling
(terms containing $\Delta_{||}$ or $\Delta_{\perp}$ in Eqs. (\ref{fulleqswithpinning1})--(\ref{fulleqswithpinning4}))
is not sufficiently large to unpin the walls the walls absorb the momentum leading
to oscillations. The threshold current is the current at which depinning occurs.
The depinning process is clearly aided by the force from the separation of the domain walls and once the
walls clear the pinning potentials they start to travel together with the difference in phases eventually vanishing.
In this limit the velocity decreases as discussed in Sec. \ref{terminal}.
We calculate domain wall motion in the presence of pinning potentials from the velocity in this limit.

\subsubsection{Weak non-adiabatic force ($\beta < \alpha$)}

\label{sec:weak}
We focus first on regime of weak force from the non-adiabatic torque $\beta$.
We set $k_{1,2}=0.1$, $\beta=0.005$ and the damping term $\alpha=0.01$. 
We set the potential well width $\xi=1$ which is comparable in size to the domain wall width.
We find also that the potential well width
does not significantly affect the results since the wall motion is coupled also
inside very wide potential wells.
The in-plane interlayer coupling is assumed to be antiferromagnetic $\Delta_\parallel>0$ and we first neglect
the perpendicular component  in the calculations setting $\Delta_\perp=0$. We investigate the effect of the perpendicular component later.

\begin{figure}
\includegraphics[width=\columnwidth]{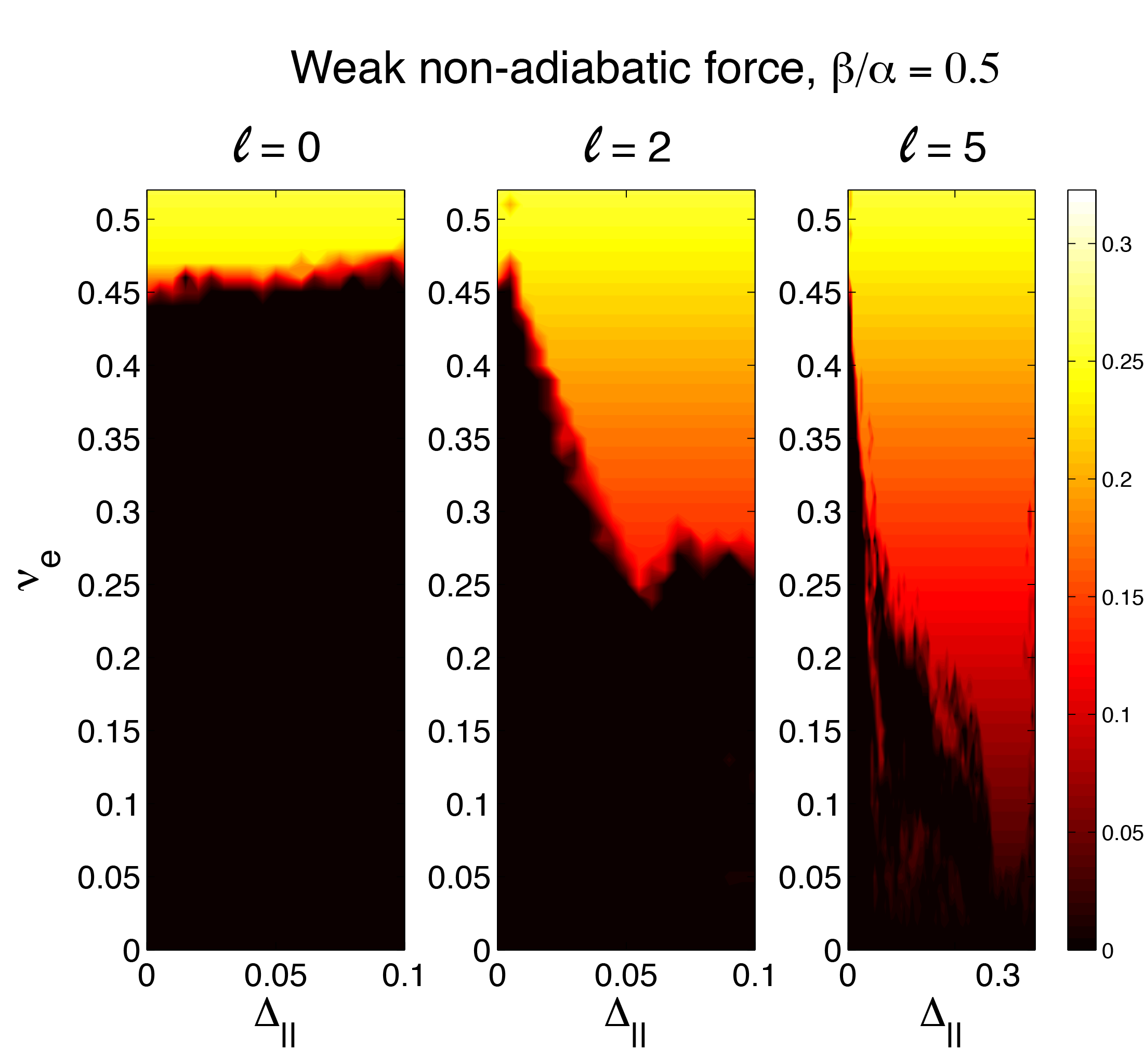}
\caption{Averaged terminal domain wall velocity (given by the color bar) in double-layer systems under weak non-adiabatic force ($\beta/\alpha=0.5$).
The velocity is calculated as a function of antiferromagnetic in-plane interlayer coupling $\Delta_{||}$ and velocity of driving electrons $\nu_e$.
The thickness of the layers are  $\mu_1=1$ and $\mu_2=1/2$. The velocities are calculated for different distances between the pinning potentials $\ell$.
The interlayer coupling helps unpin the domain walls and lowers the threshold current at $\ell>0$. Note that the range for $\Delta_{||}$ is
larger at $\ell=5$. \label{dwmobility}}
\end{figure}

Figure \ref{dwmobility} shows terminal domain wall velocity as a function of velocity of driving electrons $\nu_e$ and
strength of the in-plane interlayer coupling $\Delta$ at different distances between the pinning potential sites $\ell$.
The velocity is calculated for asymmetric domain wall configurations
($\mu_1=1$ and $\mu_2=1/2$).

We find that the threshold current for domain wall motion decreases rapidly with increasing antiferromagnetic interlayer coupling
at finite distance between the pinning potentials. The threshold current is lowered at large distances between the pinning potential sites.
Assuming that impurities can be modeled using pinning potentials with a random distribution and no correlations between
the layers our results mean that interlayer coupling make coupled domain walls in disordered systems much easier to depin with a current.

Our result can be qualitatively explained assuming that the separation between the walls is large and the wall in the first layer
is outside of the pinning potential,
i.e.  $Z_1 \gg \xi$  and $w'(Z_-) \approx 2\,\mbox{\rm sgn}(Z_-)$ (see Fig. \ref{FIGuw}).
We further assume that the angles $\phi_\pm$ change slowly and
that the domain walls are separated so that $Z_1 >Z_2$. 
Substracting Eq. (\ref{fulleqswithpinning4}) from  (\ref{fulleqswithpinning2}) then gives
\begin{equation}
{\alpha \over \lambda}\dot{Z}_2 =  k_2Z_2-{\beta \over \lambda} \nu_e - \mu_2 \Delta_\parallel .
\label{phiminuseq}
\end{equation}
This shows that the coupling exerts a force on the domain wall which is proportional to $\Delta_\parallel$.
The threshold current due to the extrinsic pinning potential is determined by the
condition for vanishing of the total force.
The wall is depinned when the force, the right-hand side of Eq. (\ref{phiminuseq}), vanishes at the highest pinning potential strength 
\begin{align}
k_2\xi= \frac{\beta}{\lambda} \nu_e + \mu_2\Delta_\parallel .
\label{eq:vanish}
\end{align}
Threshold value of the velocity of the driving electrons $\nu_e$ is reduced in the presence of the coupling $\Delta_\parallel$.
We note that $w'(Z_-)$ increases rapidly as a function of distance between the walls and saturates for large distances.
Therefore even a small displacement for one of the walls induces a force on the coupled wall and decreases the threshold current density.
As a consequence large initial domain wall separation assists the depinning process slightly and smooths out the sharp boundary between the regimes.
In this cross-over regime some initial domain wall configurations lead to depinning of the walls.
Finite temperature in experiments may therefore assist the depinning process.
In the case of strong pinning potentials the threshold current has only a weak dependence on the pinning potential strength.\cite{TKS_PR08}
We find that even in this regime the interlayer coupling decreases the threshold current.

\subsubsection{Strong non-adiabatic force ($\beta \gg \alpha$)}

\begin{figure}
\includegraphics[width=\columnwidth]{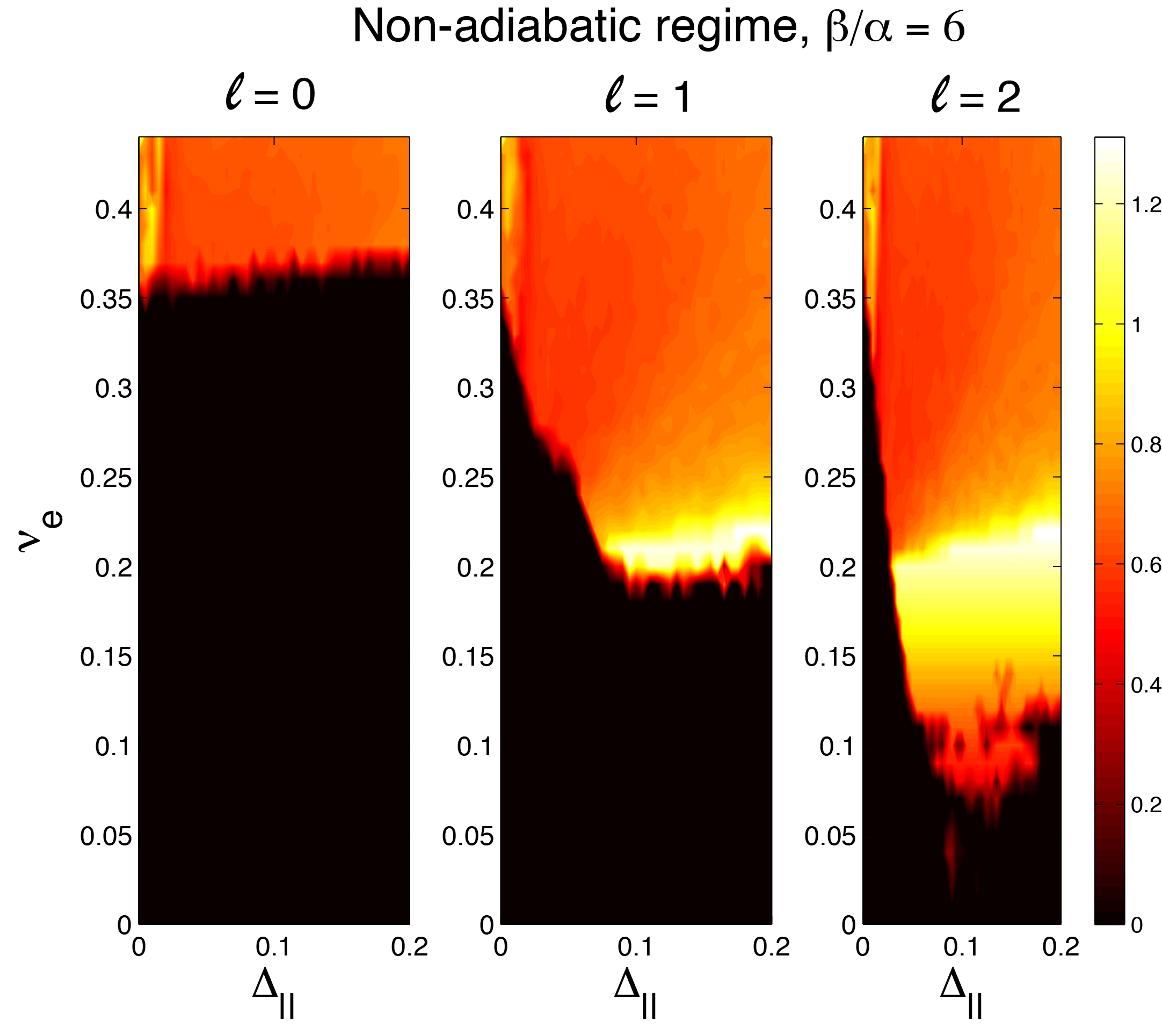}
\caption{Averaged terminal domain wall velocity (given by the color bar) in double-layer
systems in the regime of strong non-adiabatic driving ($\beta/\alpha=6$) and different distances between the pinning potential sites $\ell$.
In this regime unpinned domain wall velocity is proportional to $\beta/\alpha$ until the point of Walker breakdown at $\nu_e \simeq 0.2$.
In-plane interlayer coupling is antiferromagnetic ($\Delta=\Delta_{||} > 0$) and
improves domain wall mobility at finite $\ell$. The layers have unequal thickness  $\mu_1=1, \mu_2=1/2$ and $\alpha=0.01$.
\label{dwhighbetamobilityl0}}
\end{figure}

Next we consider the regime where $\beta$ is large and the non-adiabatic torque predominates.
The domain wall motion is then driven by the force exerted by this torque.
Threshold current density is low and the terminal domain wall velocity above the threshold current is proportional to $\beta/\alpha$ (Ref. \onlinecite{Thiaville05}).
In this regime mobility is high and the threshold current depends on the interlayer coupling and the distance between the pinning potential sites.
Figure \ref{dwhighbetamobilityl0} shows terminal domain wall velocity when the pinning potentials are located at different positions with respect to each other ($\ell=0,1,2$).
We find that the non-adiabatic force from the interlayer coupling drives the walls and the effect of the extrinsic
pinning is greatly reduced at finite $\ell$. In this regime the threshold current is strongly reduced by even a weak interlayer coupling.

At high driving currents ($\nu_e>0.2$) and weak pinning strength the domain wall mobility is reduced due to a mechanism
which is analogous to the Walker breakdown in magnetic fields. \cite{Schryer74,Hubert00}
The combination of strong non-adiabatic driving and the interlayer coupling increases domain wall mobility significantly at finite $\ell$.
We see a factor of 5 improvement in the threshold current at the coupling strength $\Delta_{\parallel}=0.1$.
Otherwise, the behaviour is similar to the case of weak non-adiabatic driving force and consistent with the analytical calculation in Sec. \ref{sec:weak}.

\begin{figure}
\includegraphics[width=\columnwidth]{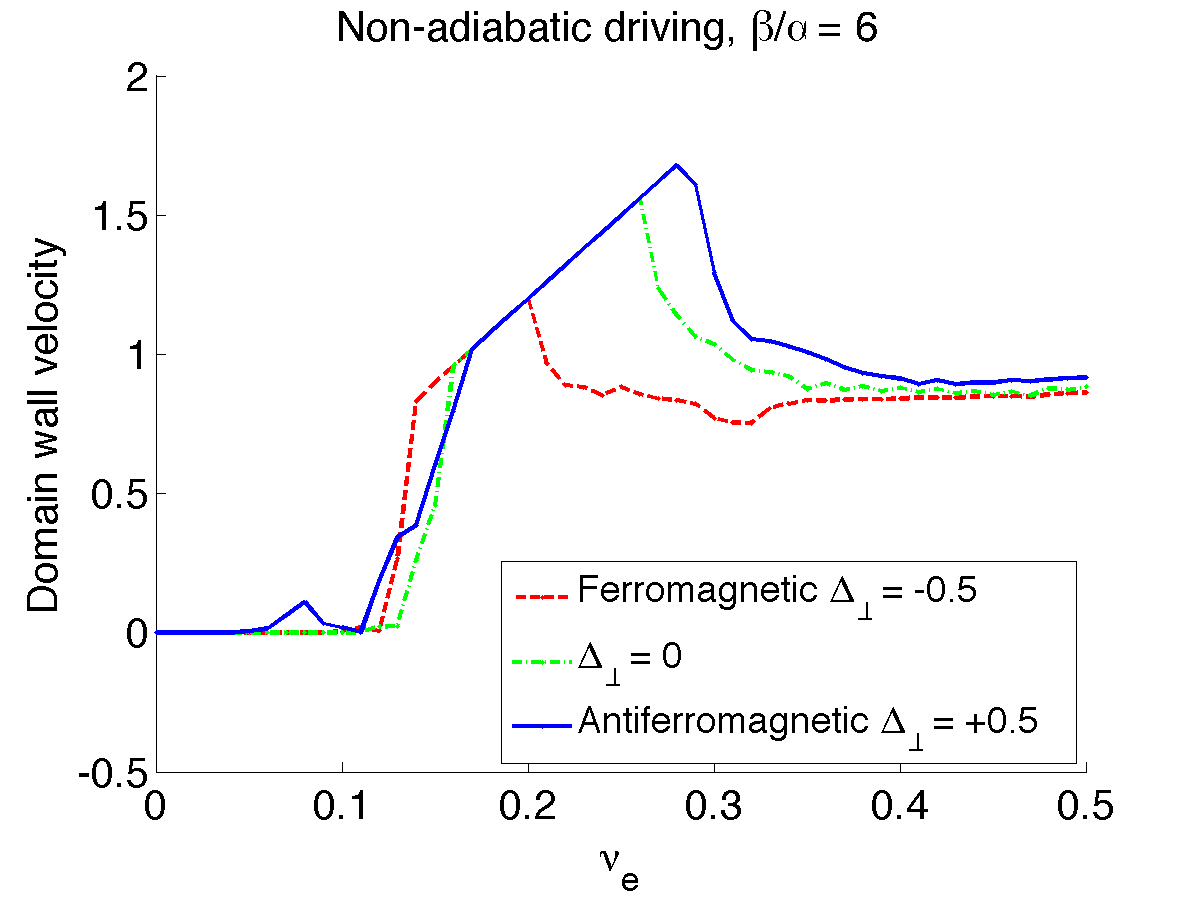}
\caption{Averaged terminal domain wall velocity in double-layer systems in the non-adiabatic driving regime
with fixed in-plane coupling strength $\Delta_{\parallel}= +0.5$ and ferromagnetic ($\Delta_{\perp}=-0.5$),
vanishing ($\Delta_{\perp}= 0$), and antiferromagnetic ($\Delta_{\perp}=+0.5$) out-of-plane component, respectively.
Layer thickness $\mu_1=1$ and $\mu_2=0.5$. Distance between the pinning potential sites $\ell=2$.
\label{dw14mobility}}
\end{figure}

So far we have neglected the out-of-plane component of the interlayer coupling  $\Delta_\perp $. 
The out-of-plane component is affected by the demagnetization field and therefore it can differ from the in-plane coupling in experiments.
However, calculations at fixed in-plane coupling strength $\Delta_{||}=0.5$ with ferromagnetic and antiferromagnetic out-of-plane coupling strength ($\Delta_{\perp}=+0.5$
and $\Delta_{\perp}=-0.5$, respectively)
show little effect on the threshold current (Fig. \ref{dw14mobility}).
Antiferromagnetic out-of-plane coupling gives the highest domain wall velocity close to the regime  of Walker breakdown.
We find also that the potential well width $\xi$
does not significantly affect the results since the wall motion is coupled also
inside very wide potential wells.

\subsubsection{Evolution of metastable states}

\begin{figure}
\includegraphics[width=0.8\columnwidth]{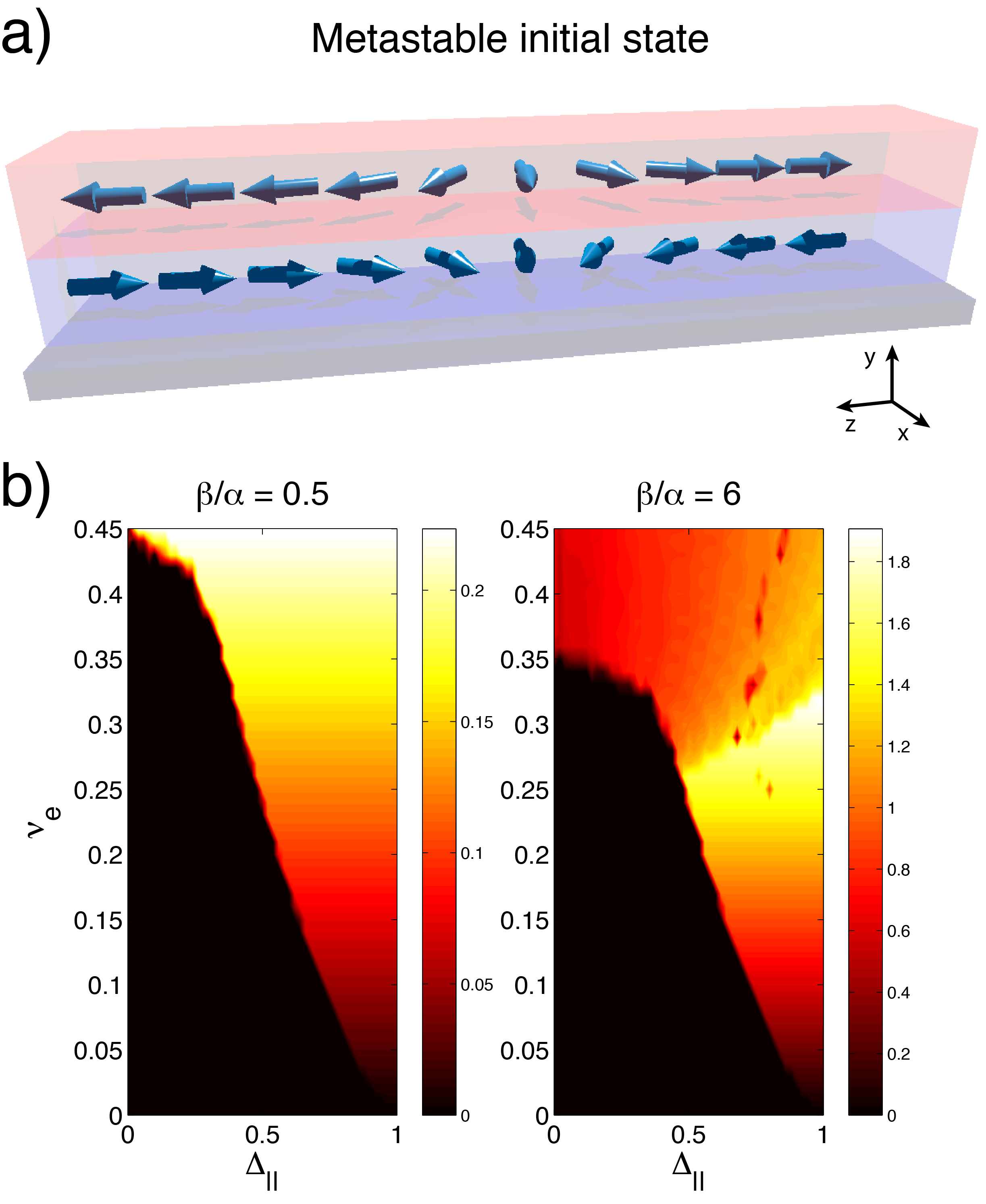}
\caption{a) A metastable initial state with parallel spin orientations at the center of the walls ($\phi_1=\phi_2=0$).
b) Averaged terminal domain wall velocity for the initial metastable states.
The in-plane interlayer coupling is antiferromagnetic $\Delta_{||} > 0$ and the out-of-plane component $\Delta_\perp=0$.
Figure shows the regime of weak non-adiabatic force ($\beta/\alpha=0.5$) and the regime of strong non-adiabatic driving ($\beta/\alpha=6$).
The threshold current decreases with increasing interlayer coupling in both cases.
The layers have unequal thickness  $\mu_1=1, \mu_2=1/2$, the pinning potential strength $k_1=k_2=0.1$ and the pinning sites are at the same position $\ell=0$.
\label{metastable-fig}}
\end{figure}

Next we study metastable states corresponding to parallel spin alignment at the center of the walls
($\phi_1=\phi_2=0$) (see Fig.  \ref{metastable-fig}a) .
Energy associated with this initial spin alignment increases with interlayer coupling and helps depin the domain walls. 
Simulations show then that the system evolves until the final states have antiparallel spin alignment. 
The threshold current decreases strongly with the increasing interlayer coupling even at $\ell=0$ in the regimes of both weak and strong non-adiabatic driving (Fig. \ref{metastable-fig}b).
We find an order of magnitude difference in the threshold current density when the interlayer coupling strength is large.

\subsection{Intrinsic pinning}

\begin{figure}
\includegraphics[width=\columnwidth]{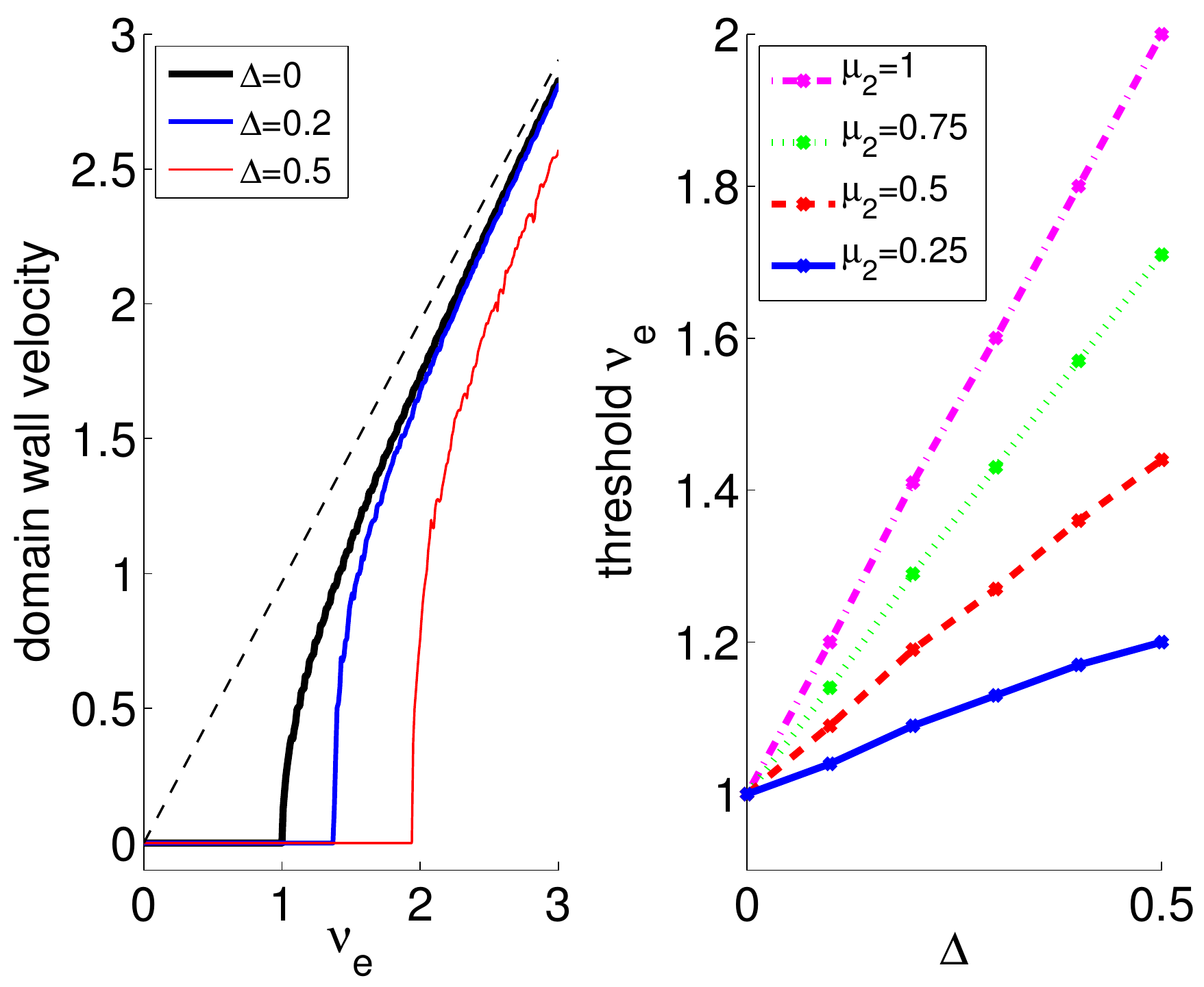}
\caption{Terminal domain wall velocity under intrinsic pinning in the absence of extrinsic pinning ($k_1=k_2=0$).
Interlayer coupling is assumed to be antiferromagnetic and isotropic $\Delta_{||}=\Delta_{\perp}=\Delta$.
The domain wall velocity as a function of the velocity of driving electrons $\nu_e$ at $\Delta=0$, $0.25$, and $0.5$ at equal layer thickness $\mu_1=\mu_2=1$  shown at left.
The threshold $\nu_e$ at different layer thicknesses $\mu_2$ fixing $\mu_1=1$ shown at right.
\label{intrinsicpinning}}
\end{figure}

\begin{figure}
\vspace{0.5cm}
\includegraphics[width=\columnwidth]{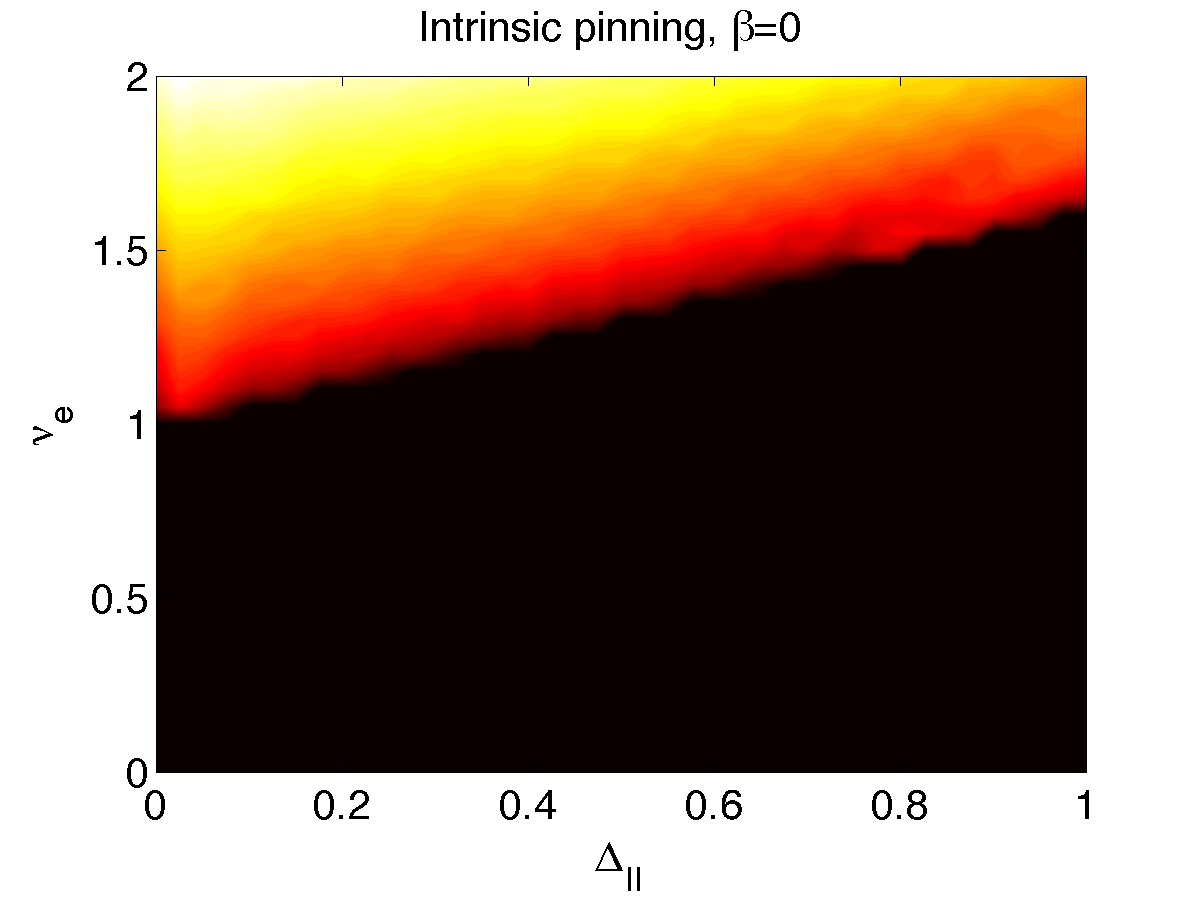}
\caption{Averaged terminal domain wall velocity under intrinsic pinning. Layers have unequal thickness  $\mu_1=1, \mu_2=1/2$.
In the absence of the non-adiabatic torque the interlayer coupling moderately increases the threshold current needed to move the domain walls.
\label{intrinsicpinninga}}
\end{figure}

In the case of adiabatic driving ($\beta \to 0$) the domain wall dynamics is driven purely by the spin transfer torque.
The intrinsic pinning effects dominate domain wall dynamics over extrinsic effects. \cite{TK04}
Intrinsic pinning is caused by the hard-axis magnetic anisotropy.
We investigate here  whether the interlayer coupling modifies the intrinsic pinning within our model at $\beta=0$ assuming that there are no extrinsic
pinning potentials $k_{1,2}=0$.
Figure \ref{intrinsicpinning} shows the domain wall terminal velocity
as a function of velocity of driving electrons at zero and finite interlayer coupling $\Delta$.
In the absence of interlayer coupling a high threshold current is needed to move the domain wall as discussed in Ref. \onlinecite{TKS_PR08}.
The threshold current is attributed to the fact that a domain wall can absorb spin torque and deform instead of the torque
setting the domain wall into motion.
Calculations indicate that intrinsic pinning is stronger in the presence of interlayer coupling  (Fig. \ref{intrinsicpinninga}).
The domain wall motion stalls after an initial boost from the spin torque of electron
current and this effect increases with the strength of the interlayer coupling.
The effect is largest for symmetric layers (Fig. \ref{intrinsicpinning}). This effect is due to enhancement of the effective magnetic hard-axis anisotropy.
The magnetic hard-axis anisotropy is proportional to $\nu_c$. Close to the ground state spin configuration $\phi_1=0;\; \phi_2=\pi$ 
the $\Delta_\pm$-dependent terms in the equations of motion (Eqs. (\ref{fulleq1}) and  (\ref{fulleq3})) combine with the term which 
is proportional to $\nu_c$ and therefore the anisotropy is effectively higher.

\section{Discussions}

In our theory the force from the interlayer coupling facilitates significantly domain wall motion in the regime where extrinsic pinning effects
dominate. Even a low interlayer coupling improves mobility and in the limit of
high interlayer coupling our theory predicts a much lower threshold current for the domain walls pinned at random impurity sites.
A domain wall separation which is of the order of the wall width is large enough to significantly improve domain wall mobility. 
This effect is further expected to be enhanced at elevated temperatures due to thermal fluctuations in the domain wall positions.
In contrast to the extrinsic pinning regime, the interlayer coupling enhances effectively the hard-axis isotropy giving rise to no
mobility improvement in the regime where intrinsic pinning effects dominate.

We have shown theoretically that interlayer coupling improves domain wall mobility in correlated bilayer systems.
The interlayer coupling greatly reduces the effective pinning potential depth when the pinning potentials are
uncorrelated in the two layers. Bilayer systems are thus promising candidates for realization of efficient
domain wall control with low current densities.

\section*{Acknowledgments}
We thank Serban Lepadatu for valuable comments and discussions.
This work was supported by a Grant-in-Aid for Scientific Research (C) (Grant Nos. 25400344 (G. T.) and 26390014 (H. S.))
from Japan Society for the Promotion of Science, UK-Japanese Collaboration on Current-Driven Domain Wall Dynamics from JST,
and the \mbox{EPSRC} through grant number EP/J000337/1.

\appendix


\bibliography{gt,14}

\end{document}